\documentclass[%
 aip,
 amsmath,amssymb,
 reprint,%
]{revtex4-1}

\usepackage{graphicx}
\usepackage{dcolumn}
\usepackage{bm}

\usepackage[utf8]{inputenc}
\usepackage[T1]{fontenc}
\usepackage{mathptmx}
\usepackage[amssymb]{SIunits}
\usepackage{hyphenat}
\usepackage[bookmarks=false]{hyperref}
\usepackage[capitalise]{cleveref}
\usepackage{floatrow}
\usepackage{graphicx}
\usepackage[caption=false]{subfig}
\usepackage{comment}

\begin{document}

\preprint{AIP/123-QED}

\title[Direct measurement of the recovery time of superconducting nanowire single-photon detectors]{Direct measurement of the recovery time of superconducting nanowire single-photon detectors}


\author{Claire Autebert}
\affiliation{Group of Applied Physics, University of Geneva, CH-1211 Geneva, Switzerland}
\author{Ga\"etan Gras}
\email{gaetan.gras@idquantique.com}
\affiliation{Group of Applied Physics, University of Geneva, CH-1211 Geneva, Switzerland}
\affiliation{ID Quantique SA, CH-1227 Carouge, Switzerland}
\author{Emna Amri}
\affiliation{Group of Applied Physics, University of Geneva, CH-1211 Geneva, Switzerland}
\affiliation{ID Quantique SA, CH-1227 Carouge, Switzerland}
\author{Matthieu Perrenoud}
\affiliation{Group of Applied Physics, University of Geneva, CH-1211 Geneva, Switzerland}
\author{Misael Caloz}
\affiliation{Group of Applied Physics, University of Geneva, CH-1211 Geneva, Switzerland}
\author{Hugo Zbinden}
\affiliation{Group of Applied Physics, University of Geneva, CH-1211 Geneva, Switzerland}
\author{F\'elix~Bussi\`eres}
\affiliation{Group of Applied Physics, University of Geneva, CH-1211 Geneva, Switzerland}
\affiliation{ID Quantique SA, CH-1227 Carouge, Switzerland}

\date{\today}

\begin{abstract}
One of the key properties of single-photon detectors is their recovery time, i.e.~the time required for the detector to recover its nominal efficiency. In the case of superconducting nanowire single-photon detectors (SNSPDs), which can feature extremely short recovery times in free-running mode, a precise characterisation of this recovery time and its time dynamics is essential for many quantum optics or quantum communication experiments. We introduce a fast and simple method to characterise precisely the recovery time of SNSPDs. It provides full information about the recovery of the efficiency in time for a single or several consecutive detections. We also show how the method can be used to gain insight into the behaviour of the bias current inside the nanowire after a detection, which allows predicting the behaviour of the detector and its efficiency in any practical experiment using these detectors.
\end{abstract}

\maketitle

%

\section{Introduction}

Single-photon detectors are a key component for optical quantum information processing.
Among the different technologies developed for single-photon detection, superconducting nanowire single-photon detectors (SNSPDs) have become the first choice of many applications showing performances orders of magnitude better than their competitors.
These nano-devices have stood out as   highly-promising detectors thanks to their high detection efficiency~\cite{2013_Marsili}, 
low dark count rate~\cite{2015_Shibata}, 
excellent time resolution~\cite{2018_Korzh, 2019_Caloz} and fast recovery~\cite{2016_Vetter}. 
Superconducting nanowire single photon detectors have already had an important impact on demanding quantum optics applications such as long-distance quantum key distribution \cite{2018_Boaron}, 
quantum networking \cite{2014_Bussieres}, optical quantum computing \cite{2018_Qiang}, device-independent quantum information processing \cite{2015_Shalm, 2016_Yin} and deep space optical communication \cite{2015_Grein}.

Depending on the application, some metrics become more important than others and can require extensive characterisation. One example is quantum key distribution (QKD), where the recovery time of SNSPDs limits the maximum rate at which it can be performed. In such a case, studying the time evolution of the SNSPD efficiency after a detection becomes important and would give us insight into the detector's behaviour, allowing the prediction of experimental performances. Obtaining accurate information is however a non-trivial task because the recovery time is intrinsically linked to the time dynamics of the bias current flowing inside the detector.

There are several methods used to characterise the recovery time of the efficiency of a SNSPD. The first one uses the output pulse delivered by the readout circuit to gain knowledge about the recovery time dynamics. However, we cannot fully trust this method since the time decay of the output voltage pulse is inevitably affected by the amplifier's bandwidth and by all other filtering and  parasitic passive components. In the best case we can only have an indirect estimation of the efficiency temporal evolution. A second method consists of extracting the recovery time behaviour from the measurement of the detection rate as a function of the incident photons rate. This method can be performed with either a continuous-wave or a pulsed laser source. The main problem with the pulsed source configuration is that we can only probe the efficiency at time stamps multiple of the pulse period which does not give full information about the continuous time dynamics. Both methods have the drawback of only providing an average efficiency per arriving photon. They can moreover be very sensitive to external parameters such as the discriminator's threshold level. Hence, using one of these measurements does not allow one to make unambiguous predictions about the behavior of the detectors in other experiments. Another method is based on measuring the auto-correlation in time between two subsequent detections when the detector is illuminated with a continuous-wave laser \cite{2017_Miki} or a pulsed laser \cite{2006_Kerman}. This method has the clear advantage over all other methods of allowing a direct observation of the recovery of the efficiency in time, and it can therefore reveal additional details (for example the presence of afterpulsing). While the implementation of this auto-correlation method is relatively simple, the acquisition time can however be very long.

In this article we introduce and demonstrate a novel method, simple in both its implementation and analysis, to fully characterise the recovery time dynamics of a single-photon detector. This method is an improvement of the autocorrelation method mentioned above, and has the advantage of a much shorter acquisition time with no need of data post-processing. We apply it to characterise the recovery time of SNSPDs under different operating conditions and for different wavelengths. We can also use it to estimate the variation of the current inside the detector after a detection, and consequently, gain insights into what happens to the bias current when two detections occur within the time period needed by the efficiency to fully recover. This method also allows us to reveal details that are otherwise difficult to observe, such as afterpulsing or oscillations in the bias current's recovery as well as predict the outcome of the count rate measurement.


\begin{figure*}[htbp]
\includegraphics[width=0.75\columnwidth]{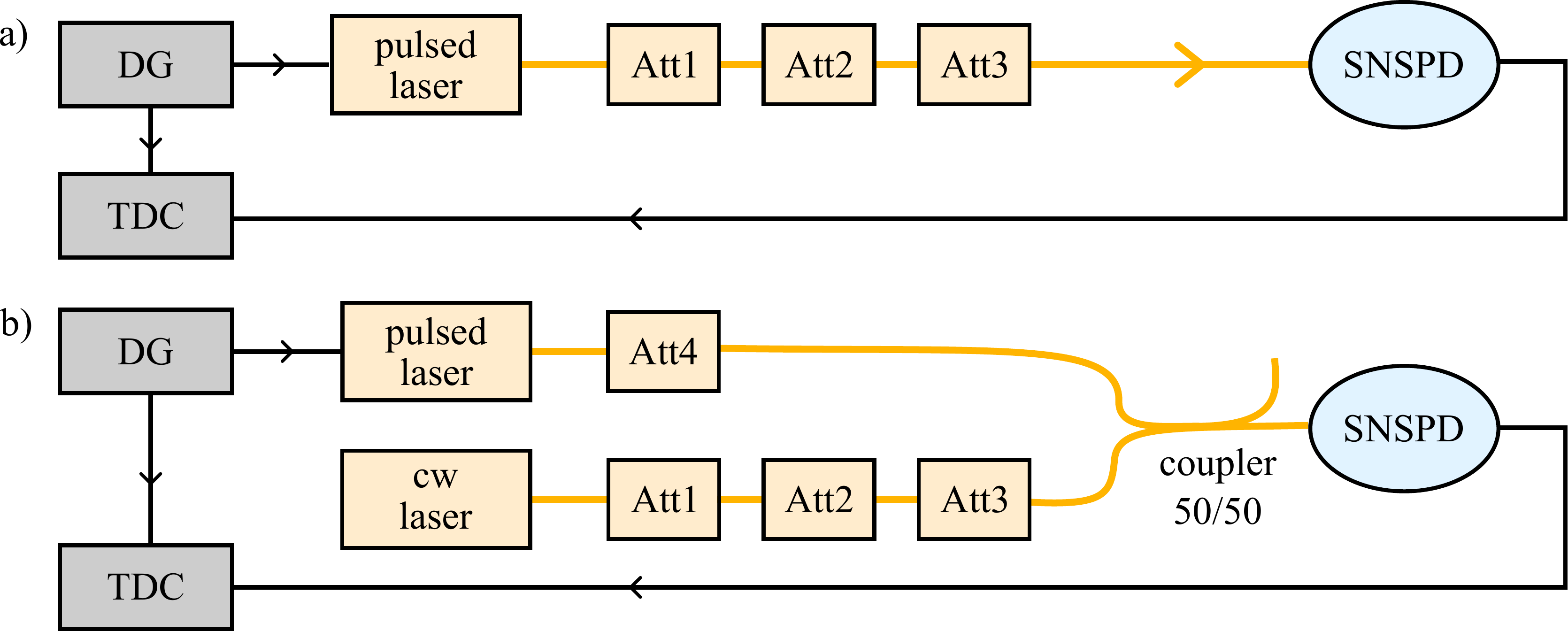}
\caption{Schematics of the experimental setups for the a) pulsed-autocorrelation method and for the b) hybrid-autocorrelation method. DG: delay generator, TDC: time-to-digital convertor, Att: attenuators.}
\label{setup}
\end{figure*}

\section{Hybrid-autocorrelation method}
To investigate the time-dependence of the detection efficiency after a first detection event, a useful tool is the normalised time autocorrelation of one detector, which is proportional to the expected probability value of having two detection events separated in time by $\Delta t$ on the same detector. For an ideal detector with a zero recovery time, the detection events occuring at times $t$ and $t + \Delta t$ are independent when illuminated with coherent light. In this case the autocorrelation will be equal to one for any value of $\Delta t$. For a detector with a non-zero recovery time, the autocorrelation function will be equal to zero at $\Delta t = 0$, and then it will recover towards one with a shape that is directly indicative of the value of the efficiency after a detection occurring at time zero. 

This method can be implemented with a continuous wave (CW) \cite{2017_Miki} or pulsed laser \cite{2006_Kerman} and it has the advantage of allowing a direct observation of the recovery of the efficiency in time. Its implementation requires a statistical analysis of the inter-arrival time between subsequent detections. A schematic of an implementation of this method with a pulsed laser is shown in Fig.~\ref{setup}a, and we use it for comparison with the novel method we introduce hereafter. A delay generator (DG) is used to generate two laser pulses with a controllable time delay between them. The triggerable laser is generating short pulses that are then attenuated down to $\approx 0.1$ photon per pulse by calibrated variable attenuators. The output signal of the detector is fed to a time-to-digital converter (TDC) that records the arrival times of the detections.  

To reconstruct the recovery of the efficiency in time after a first detection, we analyse the time stamps to estimate the probability of the second detection as a function of its delay with respect to the first one. This method can be significantly time consuming because only one given delay can be tested at once. Moreover, one needs a detection to occur in the first pulse to count the occurrences. It also requires to have the same power in both pulses and this power needs to be very stable during the whole duration of the experiment, which can be difficult to guarantee with some triggered lasers such as gain-switched laser diodes. 

Here we introduce a new method, named \textit{hybrid-autocorrelation}, that combines the pulsed and CW autocorrelation methods. The advantages of this hybrid measurement are its rapidity, flexibility in terms of wavelengths, ability to faithfully reveal the shape of the recovery of the efficiency as well as tiny features such as optical reflections in the system or even oscillations of the bias current after the detection and most importantly, it doesn't require any post-processing to extract information.  In the hybrid-autocorrelation method (Fig.~\ref{setup}b), a light pulse containing a few tens of photons is used to make the detector click with certainty at a predetermined time, which greatly reduces the total collection time needed to build the statistics. This pulse is combined on a beamsplitter with a weak but steady stream of photons (typically about $10^6$ photons/second or less) coming from an attenuated CW laser. These photons are used to induce a second detection after the one triggered by the pulsed laser, and the detection probability is proportional to the efficiency at this given time. There are no big constraints on the pulsed laser; its pulse width needs only to be much smaller than the recovery time, it does not have to be at the same wavelength as the one required for the recovery time measurement (which is determined by the CW laser) and its power and polarisation do not need to be highly stable (because its only role is to create a detection at a given time with certainty). To record the detection times we use a TDC building start-stop histograms configuration, where the start is given by the DG triggering the pulsed laser.

\section{Results}

\begin{figure}[htbp]
\includegraphics[width=0.9\columnwidth]{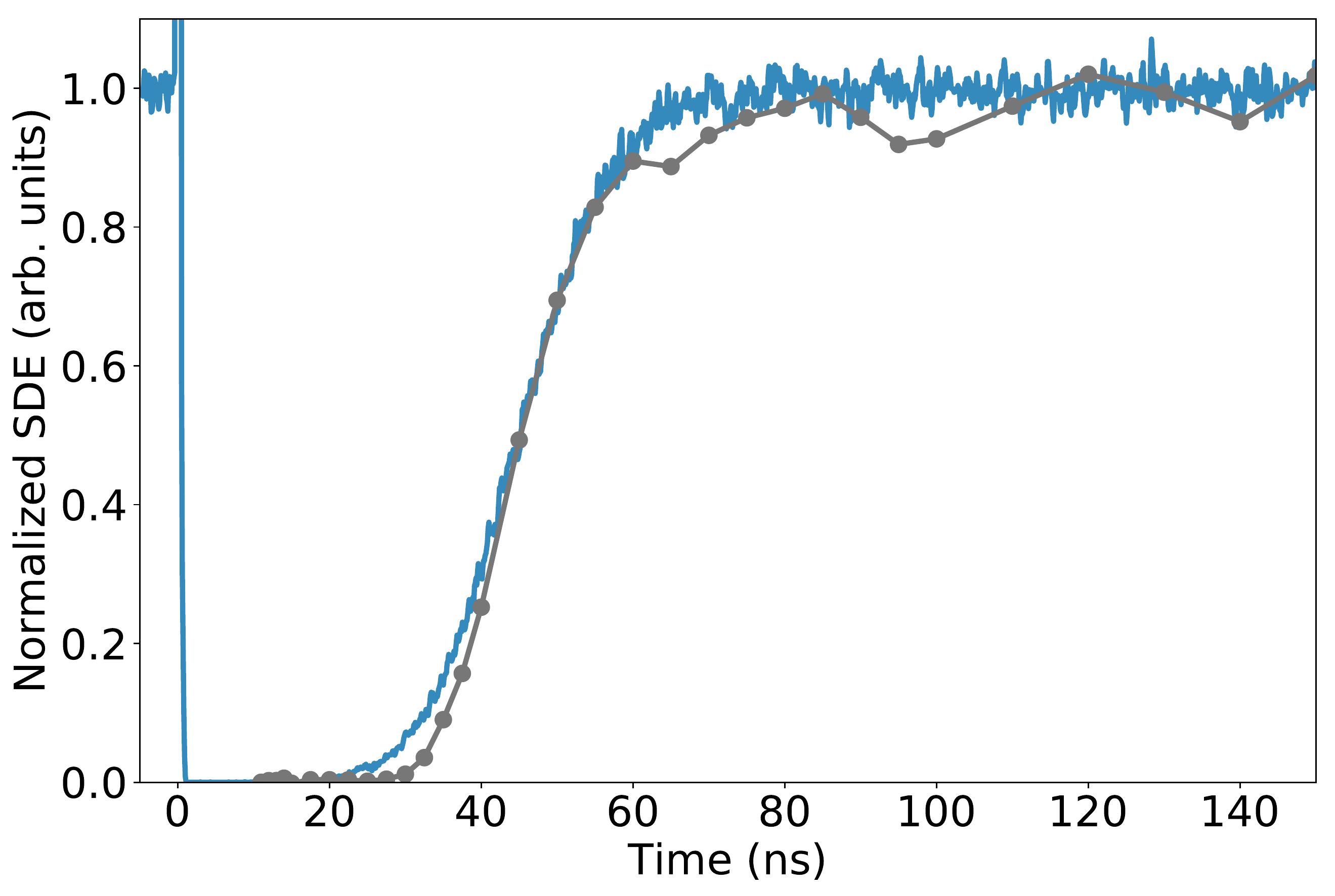}
\caption{Normalized SDE as a function of the time delay between two events for the pulsed autocorrelation method (grey points) and the hybrid-autocorrelation method (dark blue curve).}
\label{fig:autoVShybrid}
\end{figure}

We implemented the pulsed and hybrid-autocorrelation methods using a 1550~nm gain-switched pulsed laser diode with 300~ps pulse width and a CW laser at 1550~nm (for the hybrid method). We used meandered and fibre-coupled molybdenum silicide (MoSi) SNSPDs fabricated by the U.~of Geneva group \cite{2019_Caloz}. The arrival times of the detections were recorded with a TDC (ID900 from IDQ) with 100~ps-wide time bins. \Cref{fig:autoVShybrid} shows the temporal evolution of the normalised efficiency after a first detection obtained with the pulsed and hybrid-autocorrelation methods. The detector was biased very closely to the switching current $I_{\rm{SW}}$, defined as the current at which the dark counts start to rise quickly. Both  methods yielded similar results in the trend of the curves, but the pulsed autocorrelation method gave a much larger scatter in the data. This scatter is caused by the instability of the laser power over the duration of the measurement (about 6 hours). The hybrid-autocorrelation method measurement required only about one minute of acquisition time and gave the exact shape of the recovery of the efficiency. We also noticed that the detector does not show any afterpulsing effects, otherwise the normalized efficiency curve could momentarily reach values larger than one. 

\subsection{Current and wavelength dependency}
\begin{figure}[htbp]
\sidesubfloat[]{\includegraphics[width=0.9\columnwidth]{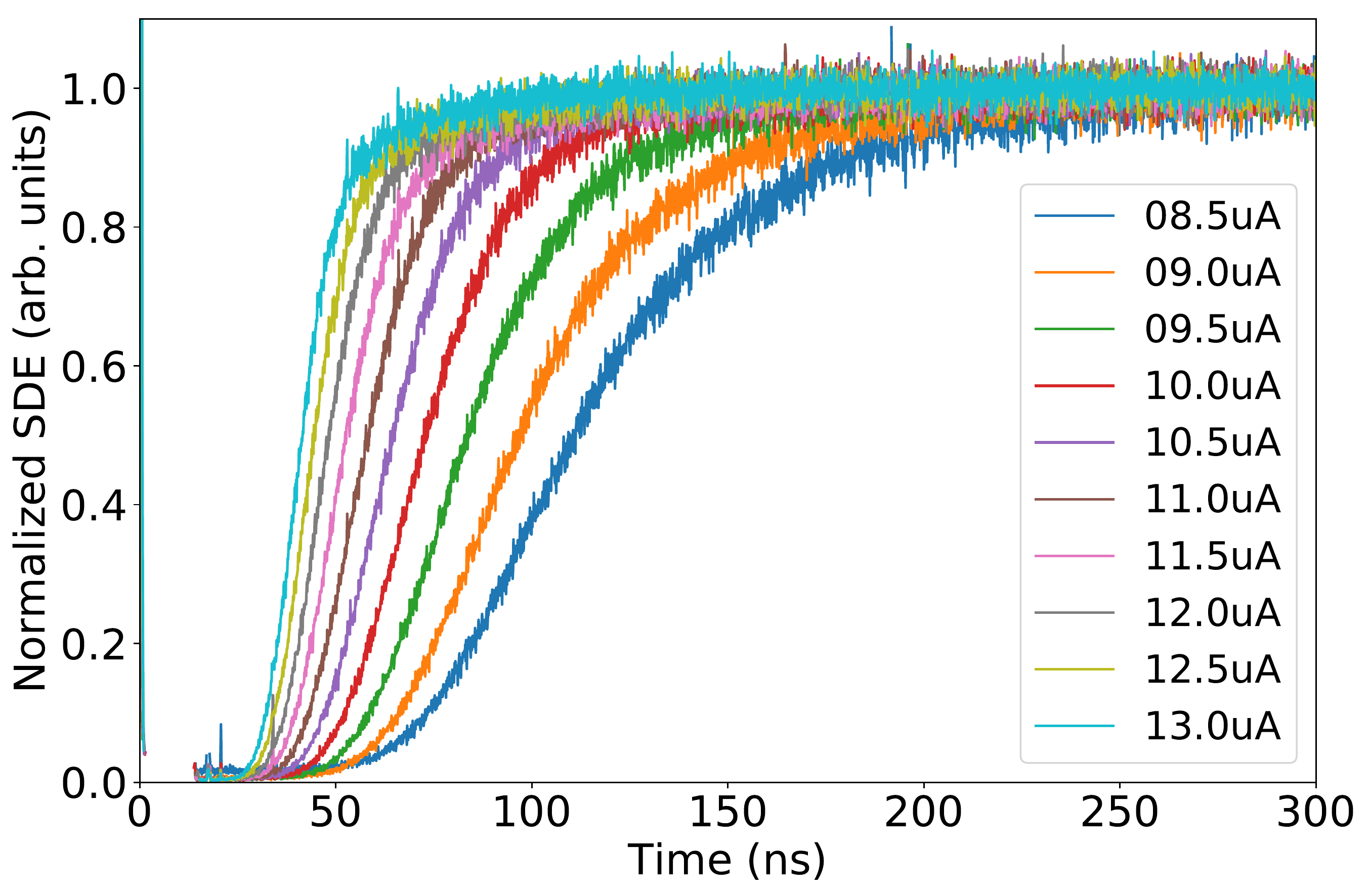}\label{sub:recVScurrent}}\\
  \sidesubfloat[]{\includegraphics[width=0.9\columnwidth]{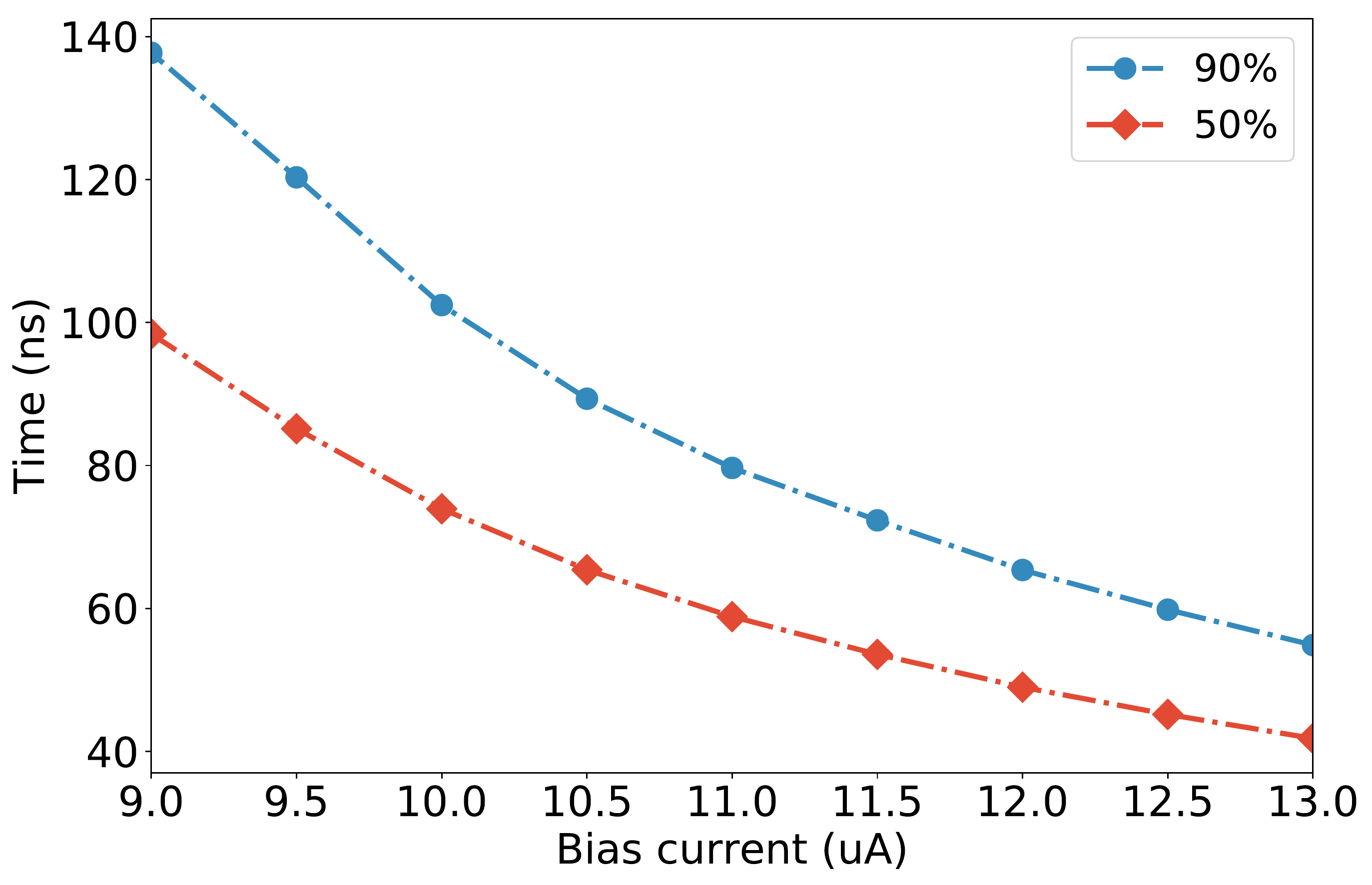}\label{sub:recVScurrentFit}}%
\caption{a) Recovery of the normalized SDE at different bias currents; b) shows the the time to recover $50\%$ (red diamonds) and $90\%$ (blue dots) of the maximum efficiency as a function of the bias current.}
\label{current}
\end{figure}

Using the hybrid-autocorrelation method we could also investigate the dependency of the recovery time on different operating conditions. First we looked at the behaviour with different bias currents. \cref{sub:recVScurrent} shows the time recovery histograms for different bias currents from $8.5~\micro\ampere$ to $13.0~\micro\ampere$, which correspond to the switching current $I_{\rm{SW}}$ of our detector. \cref{sub:recVScurrentFit} shows the time needed by the detector to recover $50\%$ (red curve) and $90\%$ (blue curve) of its maximum efficiency as a function of the bias current. The results show that the SNSPD recovery time is shorter for increasing bias current, which is expected from the shape of the efficiency curve with respect to the bias current (\cref{sub:eff}). Indeed this curve exhibits a plateau, allowing the current that is re-flowing into the nanowire after a first detection, to reach the full efficiency faster. 

\floatsetup[figure]{style=plain,subcapbesideposition=top}

\begin{figure}[htbp]
\centering\includegraphics[width=0.9\columnwidth]{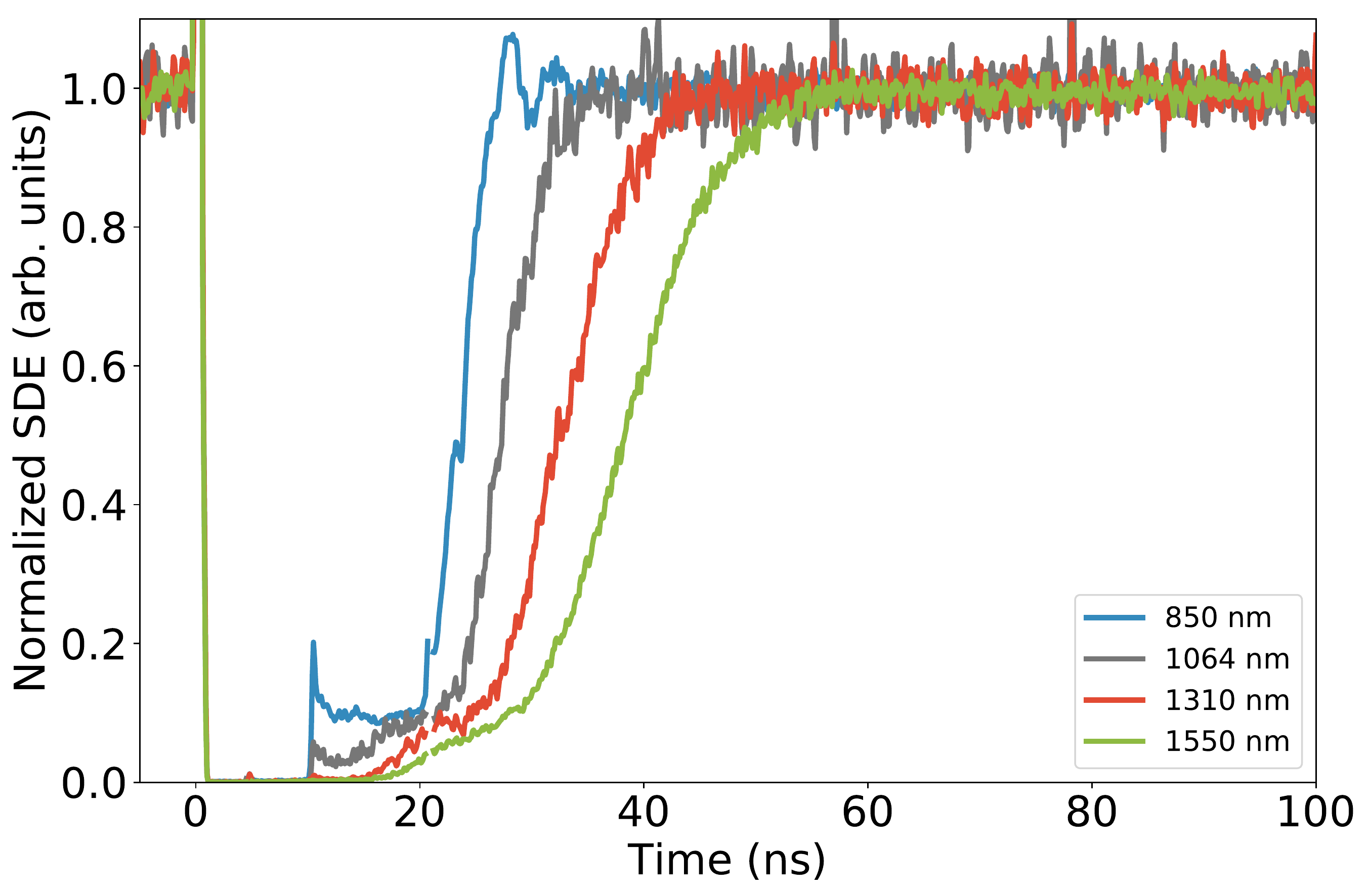}
\caption{Recovery of the normalized SDE at different wavelengths.}
\label{fig:wl}
\end{figure}

Second we vary the wavelength of the CW laser. The results are shown in \cref{fig:wl}. As expected, we can see that the lower the wavelength, the faster the recovery time. This is due to the reduction of the critical current with decreasing wavelength, while the switching current stays unchanged. This leads to an increase of the plateau length, allowing a faster recovery of the full efficiency. Interestingly, the curve at 850~nm seems to reveal some small oscillations of the efficiency around 30~ns after the trigger detection. While the origin of this small oscillation is not entirely clear (and we did not investigate this further), it nevertheless illustrates the capacity of the method to reveal some specific transient details of the efficiency recovery dynamics, or of the interplay between the voltage pulse and the discrimination circuitry.

\subsection{Current inside the SNSPD after detection}
\begin{figure}[!htbp]
  \sidesubfloat[]{\includegraphics[width=0.7\columnwidth]{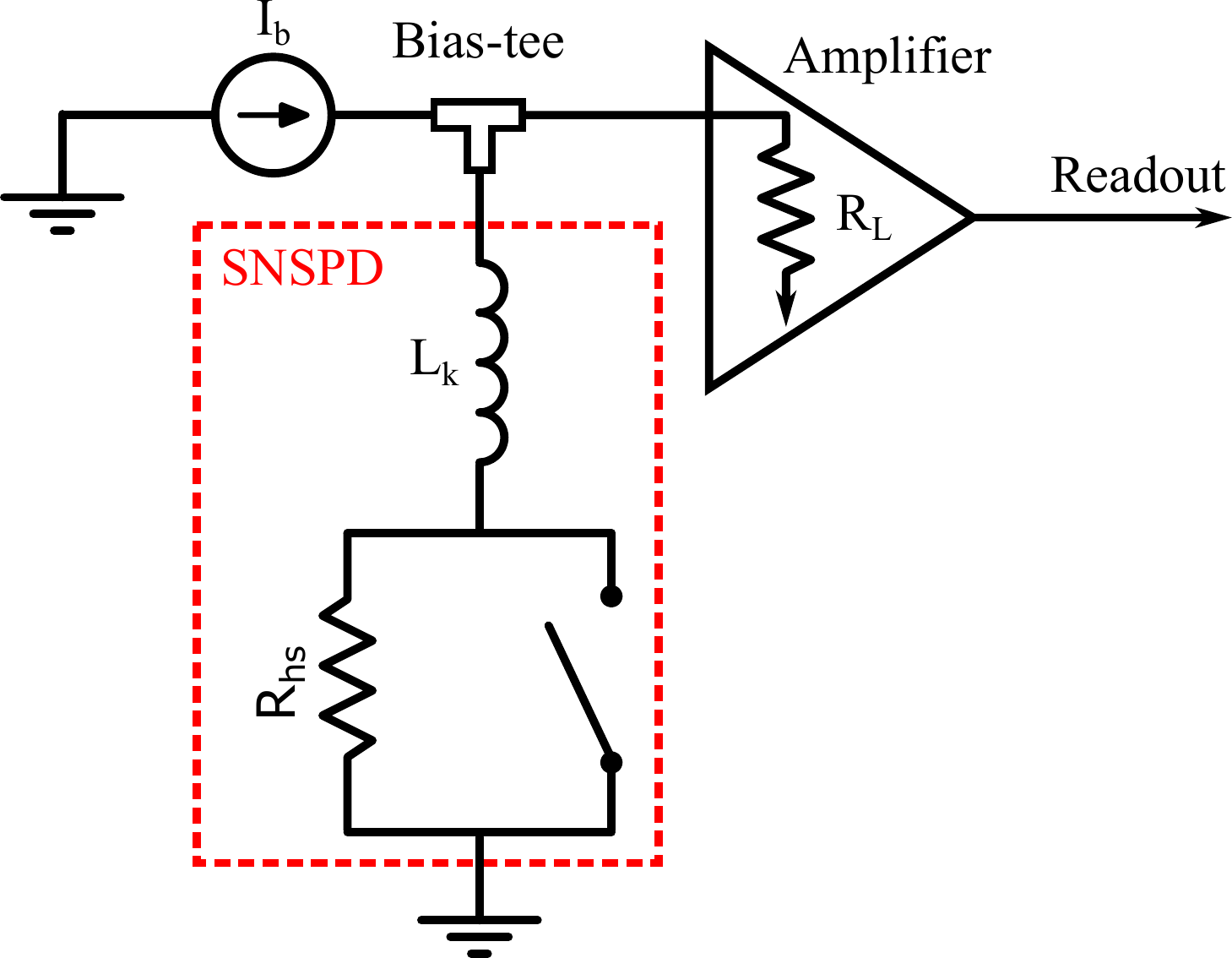}\label{sub:readout}}\\
  \sidesubfloat[]{\includegraphics[width=0.9\columnwidth]{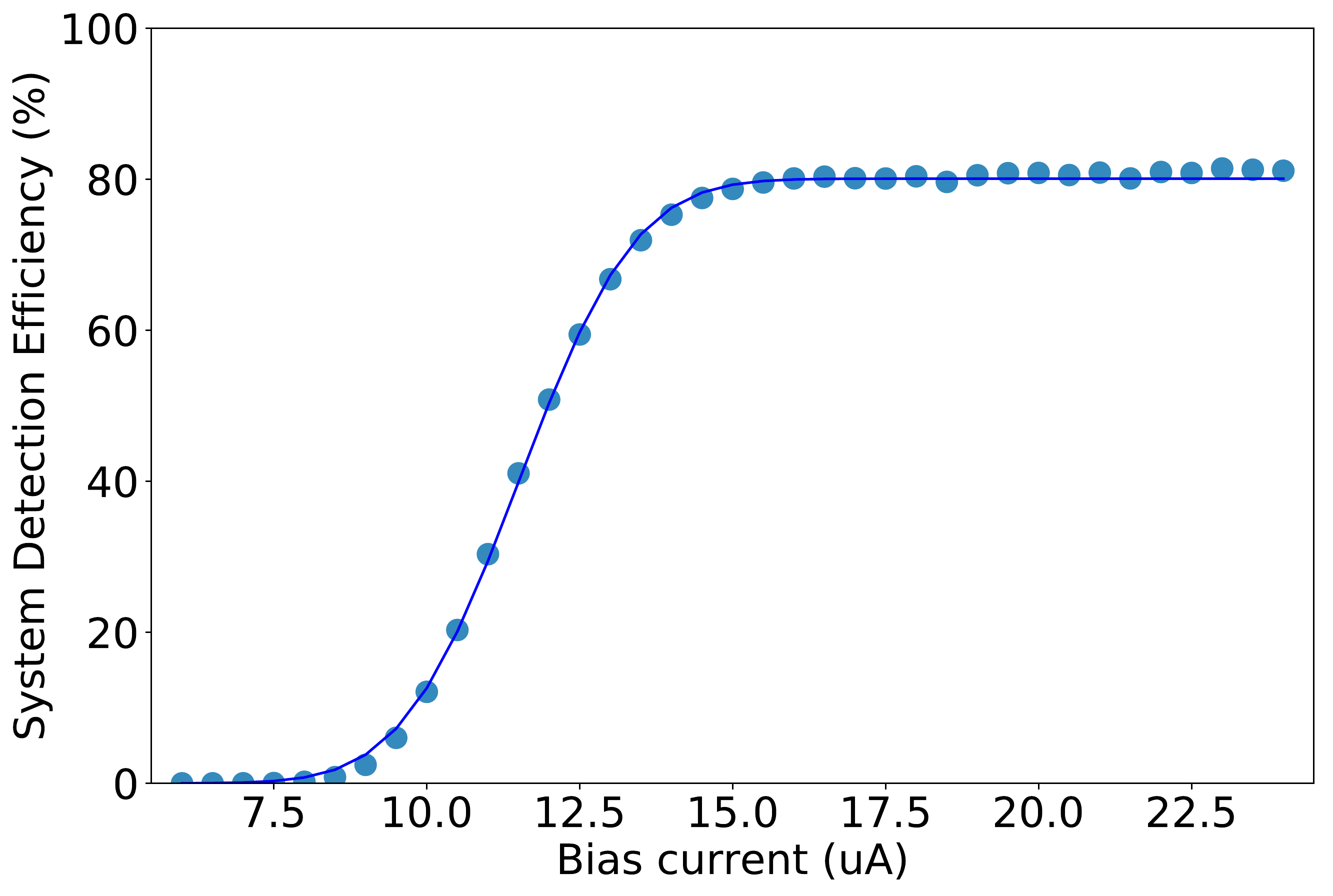}\label{sub:eff}}%
  \caption{(a) Simple equivalent electrical circuit of the detector and readout. (b) Relation between the SDE and  bias current of a typical MoSi-based SNSPD.}
  \label{fig:readout_eff}
\end{figure}

The SNSPD can be at first order modelled by an inductance $L_k$ presenting the kinetic inductance of the nanowire, serially connected to a variable resistor whose value depends on the state of the nanowire (0 if it is superconductive, $R_{\textrm{hs}} \sim 1~\kilo\ohm$ otherwise) \cite{2006_Kerman}. The bias current $I_b$ is provided by a current generator through a bias tee (see \cref{sub:readout}). When a photon is absorbed and breaks the superconductivity, it creates a local resistive region called "hotspot". The current is then deviated to the readout circuit with a time constant $\sim L_k/R_{\textrm{hs}} \sim 1~\nano\second$. Once the current has been shunted, the nanowire cools down and returns to thermal equilibrium allowing the current to return to the nanowire with a time constant of $\tau = L_k/R_L$, where $R_L = 50~\ohm$ is the typical load resistance. Note that, in practice, there may be other series resistance of a few Ohms due to the coaxial cables connecting the SNSPD to the amplifier, which might slightly increase the effective value of $R_L$, and therefore slightly decrease the value of $\tau$. Also, the amplifiers are typically capacitively coupled, which is not shown here on the drawing. The drop and the recovery of the efficiency of the SNSPD after a detection are therefore directly linked to the variation of the current and to the relation between the detection efficiency and the bias current. 

\floatsetup[figure]{style=plain,subcapbesideposition=top}

\begin{figure}[!htbp]
  \sidesubfloat[]{\includegraphics[width=0.9\columnwidth]{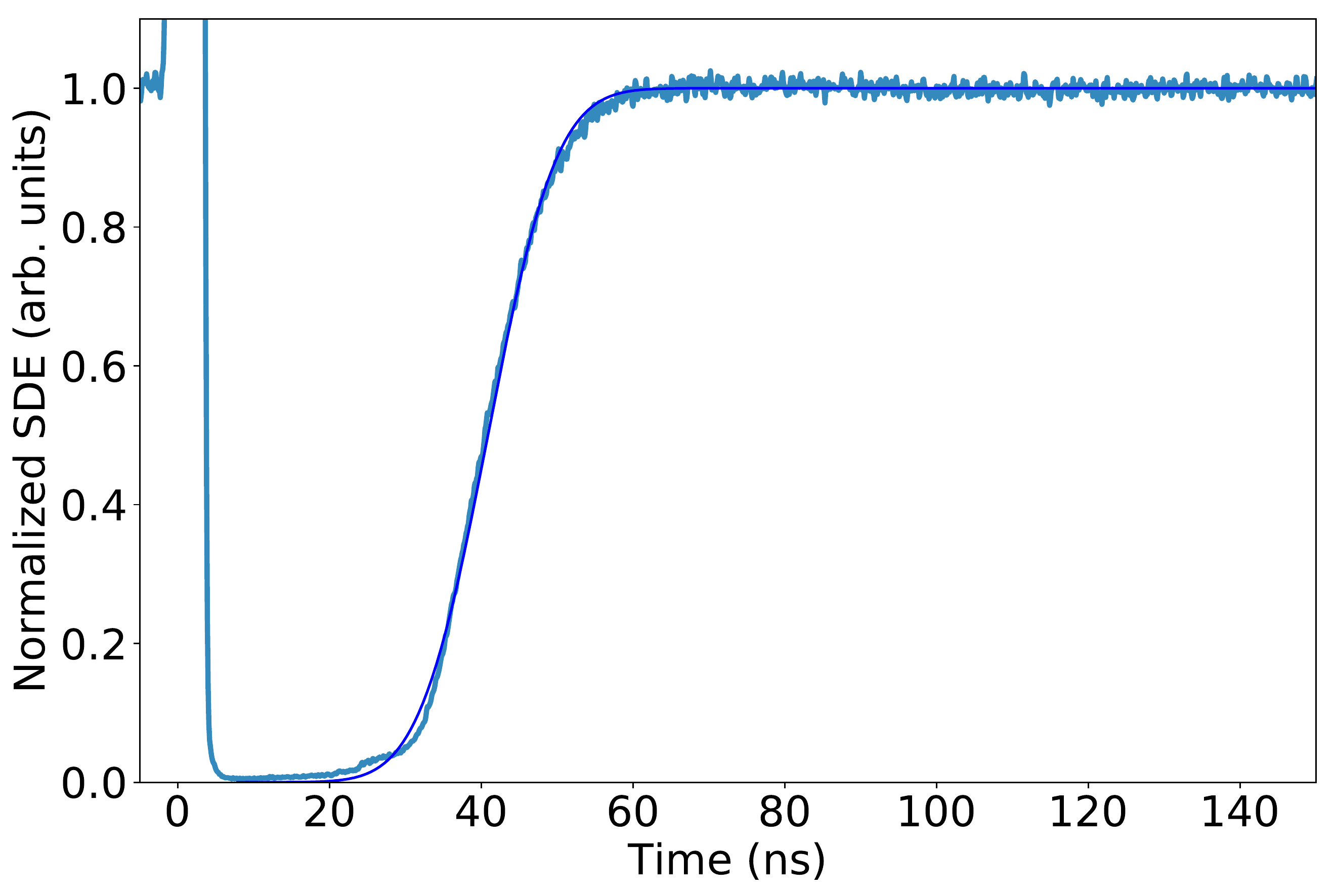}\label{sub:norm_eff}}\\
  \sidesubfloat[]{\includegraphics[width=0.9\columnwidth]{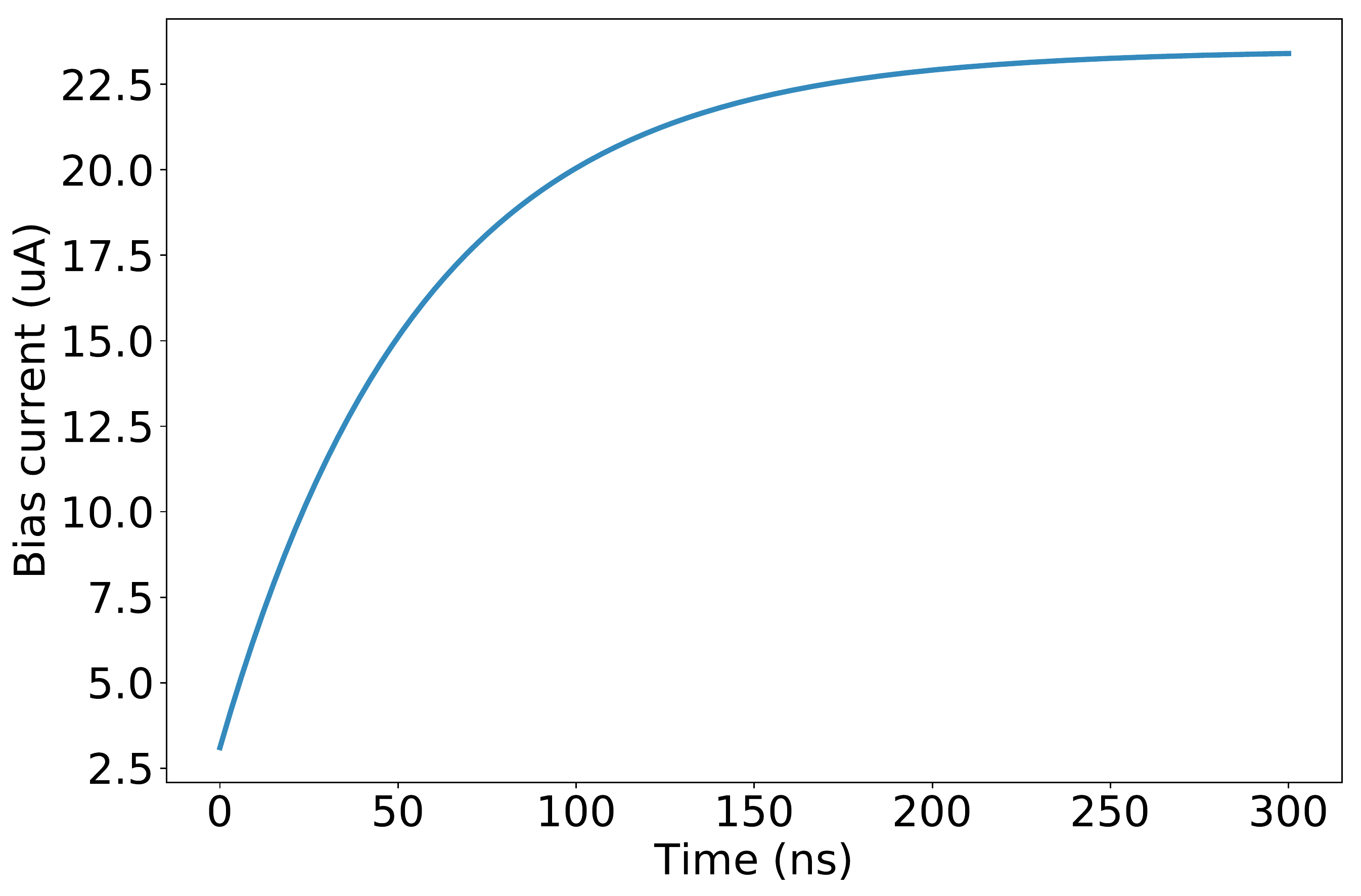}\label{sub:current}}%
  \caption{(a) Normalized efficiency as a function of time after a first detection.  (b) Reconstructed bias current of the detector as a function of time after a first detection.}
  \label{fig:currentInside}
\end{figure}

On \cref{sub:eff}, we plot the system detection efficiency as a function of the bias current of a given MoSi SNSPD, and we observe that it follows a sigmoid shape \cite{2017_Caloz}. We can therefore fit that curve using the equation
\begin{equation}
    \eta =\frac{\eta_{max}}{2}\left(1+\mbox{erf}\left(\frac{I-I_0}{\Delta I}\right)\right),
    \label{eq:eff}
\end{equation}
where $I_0$ and $\Delta I$ are parameters for the sigmoid and $\eta_{max}$ is the maximum efficiency of the detector. After a detection, the equivalent circuit of \cref{sub:readout} indicates that the current variation after a detection should be described by
\begin{equation}
    I = (I_{b}-I_{drop})\left(1-\exp\left(-\frac{t}{\tau}\right)\right)+I_{drop}.
    \label{eq:current}
\end{equation}
where $I_{b}$ is the nominal bias current of operation of the detector just before a detection, $I_{drop}$ is the current left in the nanowire immediately after a detection and $\tau$ is the time constant for the return of the current. Here, we neglect the time formation of the hotspot (and therefore the time for $I$ to go from $I_{b}$ to $I_{drop}$) as, according to the electro-thermal model of Ref.~\cite{2007_Kerman}, its lifetime is expected to be short (typically a few hundreds of ps) compared to the recovery of the current $\tau$. By fitting the curve of the efficiency versus the current with \cref{eq:eff} (\cref{sub:eff}) we can infer $I_0$ and $\Delta I$; by inserting \cref{eq:eff} in \cref{eq:current} and fitting the recovery time measurement (\cref{sub:norm_eff}) we can estimate $I_{drop}$ and $\tau$. Here, we used $I_b = 23.5~\micro\ampere$ and the best fit is obtained with $I_{drop}=3.1~\micro\ampere$ and $\tau=56~\nano\second$.  Then using both results, we can infer the value of the current in the nanowire versus time as shown on \cref{sub:current}. 
It is worth noting that this method predicts that $I_{drop}>0$. Physically, this would mean that the current did not have time to completely leave the SNSPD before it became superconductive again. This is the kind of detail that is very difficult to measure directly. Admittedly, this prediction made with our method is not direct and therefore difficult to fully confirm. We note however that we obtained values for $I_{drop}$ greater than zero for all the tested detectors. Moreover, with the values obtained for $I_{drop}$ and $\tau$, thanks to \cref{eq:eff} and \cref{eq:current} and the efficiency versus bias current and time recovery measurements, it is possible to  accurately predict the behavior of a  detector at high detection rates, as shown in \cref{section:countrate}. This gives us an increased confidence in the method proposed here. 


\floatsetup[figure]{style=plain,subcapbesideposition=top}

\begin{figure}[!htbp]
    \sidesubfloat[]{\includegraphics[width=0.9\columnwidth]{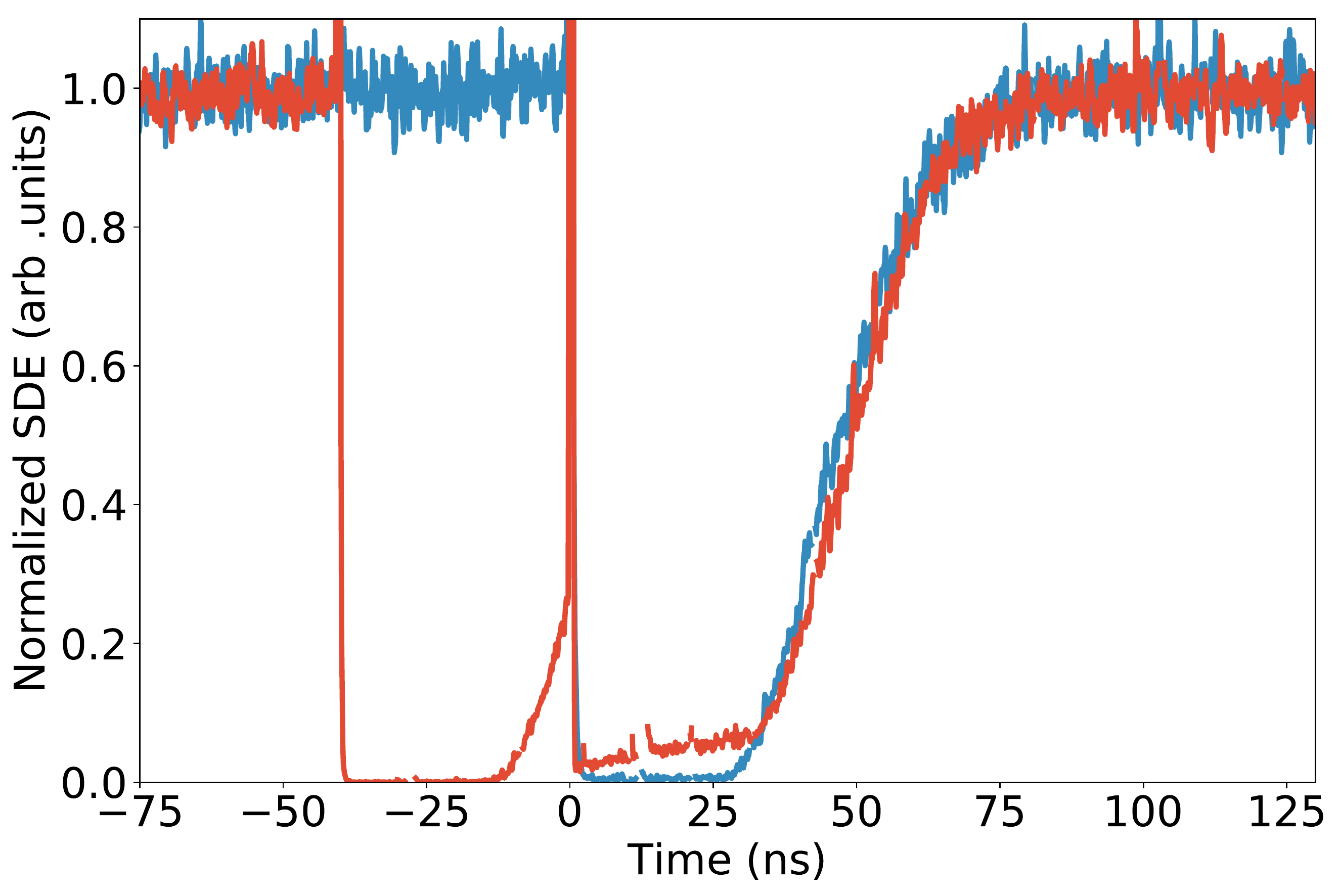}\label{sub:tp40}}\\
    \sidesubfloat[]{\includegraphics[width=0.9\columnwidth]{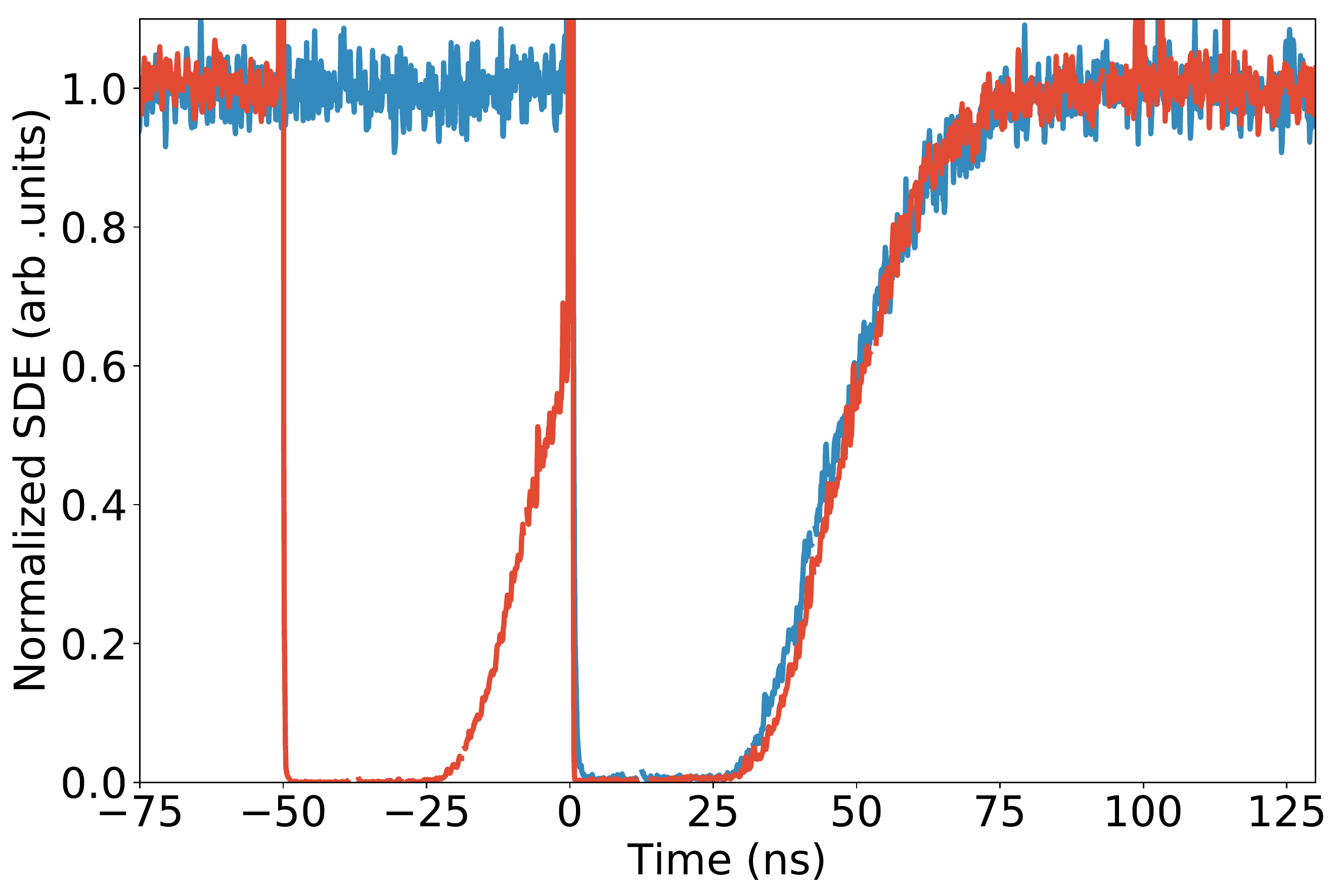}\label{sub:tp50}}\\
    \sidesubfloat[]{\includegraphics[width=0.9\columnwidth]{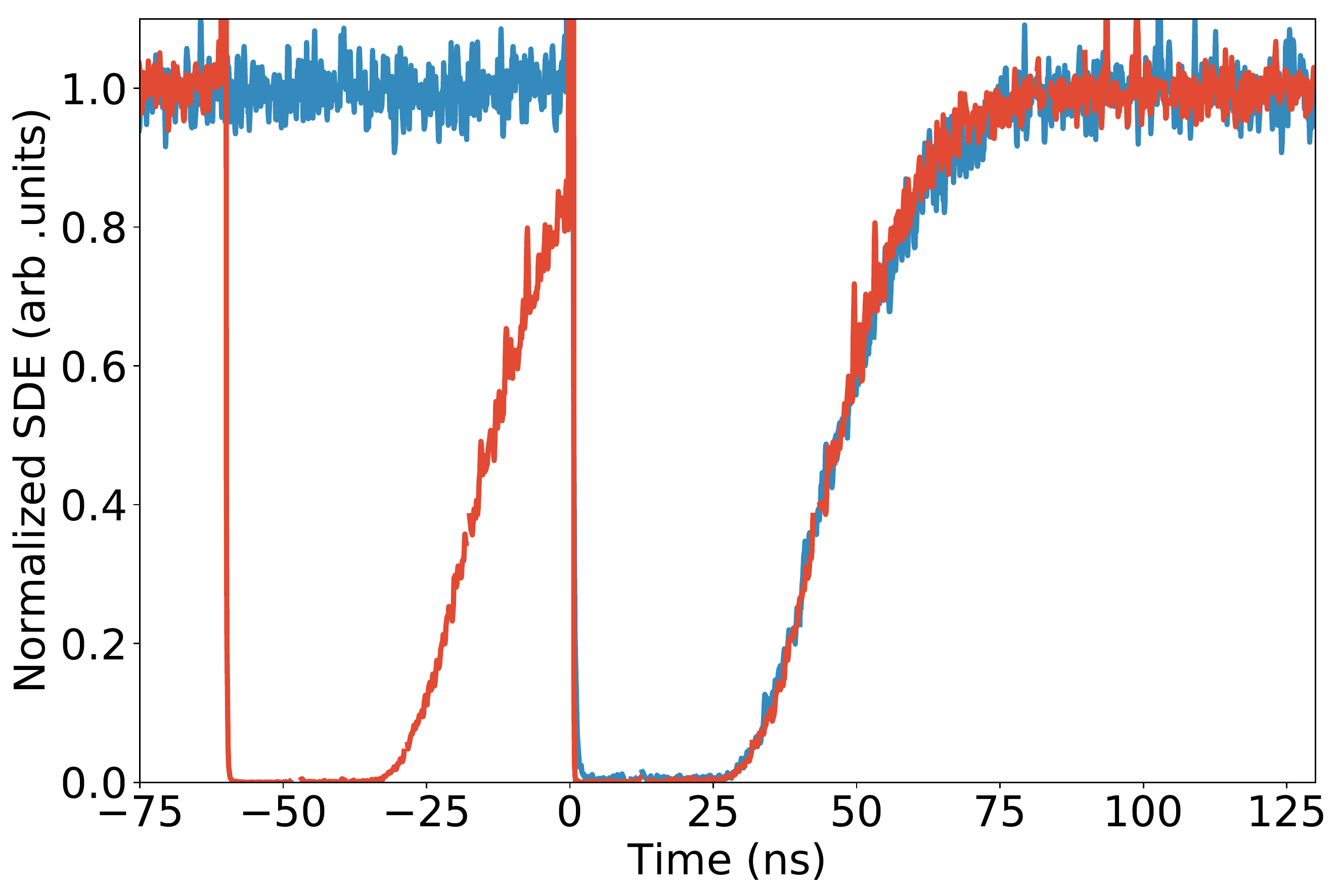}\label{sub:tp60}}%
  \caption{Recovery of the normalized SDE for one pulse (blue curve) and for two trigger pulses (red curve) with different delays between the pulses: (a) \unit{40}{\nano\second}, (b) \unit{50}{\nano\second} and (c) \unit{60}{\nano\second}.}
  \label{fig:twoPulses}
\end{figure}

When a photon strikes the nanowire and a detection occurs, the current inside the detector drops to a percentage of its original value and not to zero. An interesting measurement possible with our method consists in sending a train of pulses (here two) with varying delay between them to measure the efficiency recovery after the second detection.
With several consecutive detections, we might expect some cumulative effect with the current dropping to lower and lower values. This would lead to a longer recovery time of the detector. The results of this measurement are shown in \cref{fig:twoPulses}. The red curves correspond to the cases where two strong pulses where sent, with different time delays between them, and the blue curves correspond to the cases where only one strong pulse was sent. We can see that the shape of the autocorrelation curve for the third detection (in the case of 2 pulses) matches perfectly the one for the second detection (in the case of 1 pulse). This gives us good confidence that the current drops always to the same value. This has never been observed as clearly before despite being important for performance characterisation at high count rates. Indeed for experiment where the photons arrive with very short delays between them, it is important to know that the recovery time after any detection is the same and is not affected by the time delay between detections.

\floatsetup[figure]{style=plain,subcapbesideposition=top}

\subsection{Predicting the counting rate with a continuous wave source} \label{section:countrate}
We illustrate the predictive power of the hybrid-autocorrelation method proposed here by looking at the behaviour of SNSPDs at high counting rate, when the average time between two detections becomes comparable to the recovery time of the SNSPD. We model an experiment where the light of a continuous-wave laser is sent to the detector and the detection rate is measured as a function of the incident photons rate. To estimate the count rate versus incident photon rate from the hybrid-autocorrelation method, we run a Monte-Carlo simulation. We randomly select the time $t$ of arrival of the photon since the last detection using the exponential distribution (which gives the probabilitity distribution of time intervals between events in a Poissonian process). Thanks to the autocorrelation measurement, we know the probability of a successful event (i.e. a detection) at time $t$. In case of unsuccessful event, we look at the time $t+t'$ of arrival of the next photon. Once we have a detection, we start over. We run this until we have $N = 10~000$ detections to estimate the count rate of the detector.

\begin{figure}[!htbp]
\centering\includegraphics[width=0.9\columnwidth]{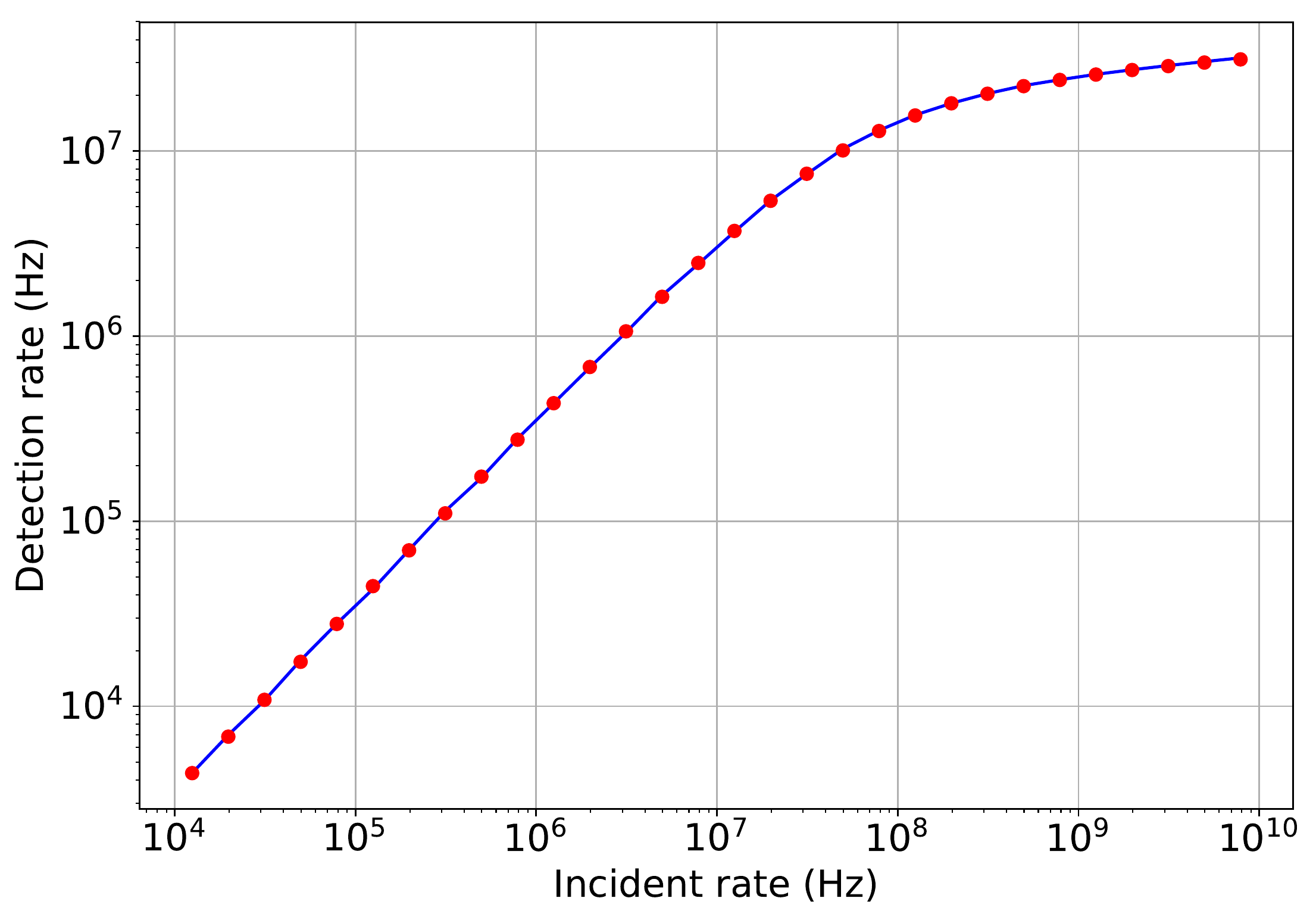}
\caption{Count rate with continuous wave laser: the red dots correspond to the count rate measurement versus incident photon rate, and the blue curve correspond to the prediction from the hybrid-autocorrelation measurement.}
\label{fig:countRate}
\end{figure}

\Cref{fig:countRate} shows, for one of the SNSPD we tested, the comparison between the experimental detection rate versus incident photon rate of the SNSPD and its prediction from the hybrid-autocorrelation measurement. 
We can see that the count rate data and the extrapolation from the autocorrelation measurement with $I_{drop}=2.9~\micro\ampere$ and $\tau=58~\nano\second$ fit very well together, giving a high trust in the model and in the predictive power of the method.

\section{Conclusion}
The method we proposed here provides a fast, simple and most importantly direct characterisation of the recovery of the efficiency of a SNSPD detector. The measurements showed that the recovery of a SNSPD is faster with larger bias current and shorter wavelengths. We demonstrated that the current through a given  detector always drop to the same non-zero value after detection even when subjected to several consecutive pulses all arriving within a fraction of the total recovery time of the SNSPD. We also showed that our method can be used to correctly predict how the detection rate of an SNSPD behaves when it becomes impeded by its recovery time. Therefore, we trust our method to allow predicting the behavior of the SNSPD in other experiments where the variation of the efficiency in time is of importance.
Finally, it is also worth noting that this method can be applied to any type of single-photon detector, and could be considered as a universal benchmarking method to measure and compare the recovery time of single-photon detectors.

\begin{acknowledgments}
This project was funded from the European Union's Horizon 2020 programme (Marie Sk\l{}odowska-Curie grant 675662).
\end{acknowledgments}

\bibliography{aipsamp}

\end{document}